# Improvement in thermoelectric properties by tailoring at In and Te site in $In_2Te_5$

*Anup V. Sanchela$^{a*}$, Ajay D. Thakur$^{b,c}$, C. V. Tomy $^a$*

$^a$ *Department of Physics, Indian Institute of Technology Bombay, Mumbai-400076, India*

$^b$ *School of Basic Sciences, Indian Institute of Technology Patna, Patna-800013, India*

$^c$ *Centre for Energy and Environment,*

*Indian Institute of Technology Patna, Patna-800013, India*

## Abstract

We study role of site substitutions at In and Te site in $In_2Te_5$ on the thermoelectric behavior. Single crystals with compositions $In_2(Te_{1-x}Se_x)_5$ ($x$ = 0, 0.05, 0.10) and $Fe_{0.05}In_{1.95}(Te_{0.90}Se_{0.10})_5$ were prepared using modified Bridgman-Stockbarger technique. Electrical and thermal transport properties of these single crystals were measured in the temperature range 6 - 395 K. A substantial decrease in thermal conductivity is observed in Fe substituted samples attributed to the enhanced phonon point-defect scattering. Marked enhancement in Seebeck coefficient *S* along with a concomitant suppression of electrical resistivity ρ is observed in Se substituted single crystals. An overall enhancement of thermoelectric figure of merit (*zT*) by a factor of 310 is observed in single crystals of $Fe_{0.05}In_{1.95}(Te_{0.90}Se_{0.10})_5$ compared to the parent $In_2Te_5$ single crystals.

$^*$Corresponding Author
 Email: phonon.sanchela@gmail.com

## 1. Introduction

The nonrenewable aspect of fossil fuels and environmental considerations such as $CO_2$ emission, global warming and ozone depletion are main issues related to sustainable development. Thereby to increase the use of renewable energy sources like solar, wind biomass is the alternative option [1, 2] and another option. Increasing the efficiency of various processes via the recovery of waste heat using thermoelectric materials is also being pursued [3, 4, 5]. They are more reliable because they have no noise or vibration as there are no mechanical moving parts. The thermoelectric efficiency can be defined by dimensionless figure of merit $zT = \frac{S^2 T}{\rho \kappa_{tot}}$ where $T$ is absolute temperature, $S$ is Seebeck coefficient, $\rho$ is electrical resistivity and $\kappa_{tot}$ is the total thermal conductivity which is sum of two components the lattice component $\kappa_l$ and electronic component $\kappa_e$. The power factor can be defined by $PF = \frac{S^2}{\rho}$. For a good thermoelectric material needs to large Seebeck coefficient, minimum electrical resistivity and thermal conductivity and maximum power factor [4, 6, 7]. Recently lot of attention has been focussed on $In_4Se_3$, $In_4Se_{3-x}$ and $In_4Te_3$ due to their high $zT$ [8-10]. For the single crystal $In_4Se_{2.35}$ reported highest $zT$ value 1.48 at 705 K in the b-c planes [10]. $In_4Se_3$ exhibit unique two dimensional crystalline sheets layered structure with weak van der Walls interactions containing charge density wave which is involved with low dimensional transport phenomena rendered lattice distortion (Peierls distortion). An enhanced phonon scattering greatly reduces thermal conductivity [10, 11, 12]. For the hot pressed polycrystalline $In_4Se_{2.35}$ and $In_4Se_{2.2}$, $zT$ values of ~1 at 698 K [9] and hot pressed $In_4Se_3$ and $In_4Te_3$ polycrystalline samples $zT$s are 0.6 at 700 K and 0.1 at 450 K along the pressing direction, respectively [8]. The energy band gaps of $In_4Se_3$ and $In_4Te_3$ materials are 0.42 eV and 0.29 eV at room temperature respectively indicating that both compounds behave as semiconductors [8]. $In_2Te_5$ is an interesting system having a layered structure [10, 12-14]. However its $zT$ is very low due to large electrical resistivity.

In this paper we report Se substituted, $In_2(Te_{1-x}Se_x)_5$ (x = 0, 0.05, 0.10) and Fe substituted, $Fe_{0.05}In_{1.95}(Te_{0.90}Se_{0.10})_5$ samples prepared by modified Bridgman-Stockbarger technique and study their thermoelectric properties. For the composition $Fe_{0.05}In_{1.95}(Te_{0.90}Se_{0.10})_5$, we have been able to improve the $zT$. This is mainly due to lower electrical resistivity and large mass difference between Fe and In that can enhanced point defect phonon scattering leading to reduced thermal conductivity [15, 16].

2. **Experimental details**

Single crystalline samples of $In_2(Te_{1-x}Se_x)_5$ x = 0, 0.05, 0.10 and $Fe_{0.05}In_{1.95}(Te_{0.90}Se_{0.10})_5$ were prepared by the modified Bridgman technique. Powder of Fe (99.99%), Rod of In (99.99%), Shots of Te (99.99%) and Se (99.99%) obtained from Alfa Aesar were mixed in stoichiometric proportion were placed into a thick quartz tube. These quartz tubes were then evacuated to $10^{-5}$ torr and heated up to 500°C at the rate of 33°C/h, kept for 24h and then slowly cooled (2°C/h) to 470°C where it was left for 24 hours after that switched off the furnace and cooled up to room temperature [13, 14]. All parallelepiped single crystals densities in the range of 3.30 to 4.02 g cm$^{-3}$, about 56.4 % to 68.7 % of the theoretical density 5.85 g cm$^{-3}$. Thermal conductivity and Seebeck coefficient were measured using the Thermal Transport Option (TTO) in a Physical Property Measurement System (PPMS) (Quantum Design Inc., USA). Resistivity and heat capacity measurement were performed using the resistivity and heat capacity option. Powder X-ray diffraction patterns were obtained by a X-ray diffraction system (Philips X'Pert Pro, Holland) using Cu-K$\alpha$ radition.

## 3. Results and discussion

Figure 1(a) shows the XRD pattern of the powdered single crystals and Rietveld refinement calculations of $In_2Te_5$. $In_2Te_5$ has a monoclinic structure with space group C12/c1. The refinement result gives the lattice parameters are a = 16.375 Å, b = 4.330 Å and c = 40.728 Å. Figure 1(b), (c), (d) and (e) shows the patterns, which is clearly exhibits that the plane of the crystal is grown along the c-direction (only [0 0 6] and [0 0 12] reflections for $In_2Te_5$ and [0 0 12] reflections for $In_2(Te_{1-x}Se_x)_5$, x = 0, 0.05, 0.10 and $Fe_{0.05}In_{1.95}(Te_{0.90}Se_{0.10})_5$ samples.

The electrical resistivity as a function of temperature is shown in Figure 2(a). The temperature range is 6 K to 395 K. The resistivity up to x = 0.10, $In_2(Te_{1-x}Se_x)_5$ samples exhibits a semiconductor behavior decreasing with temperature. With increases Se substitution decreasing the electrical resistivity but when 5% Fe substituted at In site, $Fe_{0.05}In_{1.95}(Te_{0.90}Se_{0.10})_5$ resistivity drastically reduced and increases with temperature which indicates metallic behavior. For $Fe_{0.05}In_{1.95}(Te_{0.90}Se_{0.10})_5$ sample resistivity decreases almost three order of magnitude compare to $In_2Te_5$ sample which is $\rho \approx 0.12$ Ωm and for $Fe_{0.05}In_{1.95}(Te_{0.90}Se_{0.10})_5$ for $\rho \approx 0.00039$ Ωm about 300 K.

Figure 2(b) shows temperature dependent Seebeck coefficient for all samples in the temperature range 6 K to 395 K. All samples shows positive S values indicating that majority carriers are holes (p type). For $In_2Te_5$ sample thermopower increases with increasing temperature and reaches a maximum 409 µV/K at 267 K. As the temperature further increases, S decreases to a value 163 µV/K at 390 K. In 5% and 10% Se substituted samples Seebeck coefficient is much larger than the $In_2Te_5$ compound which is about 492 µV/K, 535 µV/K at 245 K and 219 K respectively. In case of the Se substituted samples, the maxima in S-T plot shifts to lower temperature. While S value of $Fe_{0.05}In_{1.95}(Te_{0.90}Se_{0.10})_5$ exhibit lower S value over the whole temperature range as compared to parent and Se substituted samples. At 267 K, for $Fe_{0.05}In_{1.95}(Te_{0.90}Se_{0.10})_5$ S value is 2.7 times smaller than the $In_2Te_5$ when at the same temperature its electrical resistivity is three order of magnitude lower than the $In_2Te_5$ sample. The reason is 5 % doping of Fe increasing the hole type carrier concentration.

Figure 3(a) shows temperature dependent lattice thermal conductivity and in inset total and electronic thermal conductivities of $In_2(Te_{1-x}Se_x)_5$ x = 0, 0.05, 0.10 and $Fe_{0.05}In_{1.95}(Te_{0.90}Se_{0.10})_5$ samples between 6 K to 395 K. Lattice thermal conductivity of $In_2(Te_{1-x}Se_x)_5$ x = 0, 0.05, 0.10 and $Fe_{0.05}In_{1.95}(Te_{0.90}Se_{0.10})_5$ increases with increasing temperature range 6 K to 50 K follow the power law and then it reaches maximum when phonon mean free path becomes comparable to the crystal dimension [17]. After which the thermal conductivities decreases with increasing temperature at the 1/T manner [17]. The lattice thermal conductivity increases with increasing Se substitution up to x = 0.10, $In_2(Te_{1-x}Se_x)_5$ while Fe and Se substituted sample $Fe_{0.05}In_{1.95}(Te_{0.90}Se_{0.10})_5$ shows much lower lattice thermal conductivity as compare to parent compound. The room temperature lattice thermal conductivities of crystals of $In_2(Te_{1-x}Se_x)_5$ x = 0, 0.05, 0.10 and $Fe_{0.05}In_{1.95}(Te_{0.90}Se_{0.10})_5$ are ~ 0.91, 1.4, 1.5 and 0.31 W/K-m, respectively. The lattice contributions are calculated by Wiedemann-Franz law $\kappa_e = L_0T/\rho$ where $L_0 = 2.45\times10^{-8}$ $V^2/K^2$ is the Lorenz number. From $\kappa_e$ we can Find the $\kappa_l$ ($\kappa_l = \kappa - \kappa_e$). In case of $Fe_{0.05}In_{1.95}(Te_{0.90}Se_{0.10})_5$ $\kappa_{lat}$ is 0.31 W/ K-m which is about 2.9 times smaller as compared to $\kappa_{lat}$ of 0.91 W/ K-m for $In_2Te_5$ at room temperature. The results reveal that Fe substitution on In site increases the point defect phonon scattering through atomic mass difference between Fe (55.8 g/mol) and In (114.8 g/mol) thereby lowering the lattice thermal conductivities [15]. The results suggest that phonon contribution is predominant. The inset panels in Fig. 3(a) shows $\kappa$ and $\kappa_e$. It indicates that Se and Fe substituted $In_2Te_5$ compounds show enhanced electronic thermal conductivities. In case of $Fe_{0.05}In_{1.95}(Te_{0.90}Se_{0.10})_5$, $\kappa_e$ is two order of magnitude (~0.017 W/K-m) higher than $In_2Te_5$, $\kappa_e$ (~ 0.00015 W/K-m) at 390 K.

Figure 3(b) shows temperature dependent calculated phonon mean free path ($l_{ph}$). The theoretical calculated phonon mean free path can be obtained by kinetic formula, $\kappa_l = \frac{1}{3}C_{ph}v_s l_{ph}$ where $C_{ph}$ is phonon heat capacity, $v_s$ is the velocity of sound and $\kappa_l$ is the lattice thermal conductivity [17, 18]. To get the phonon heat capacity Debye Einstein fitting is used for all samples equ [18, 19].

$$C_{ph} = 9R\left(\frac{T}{\theta_D}\right)^3 \int_0^{x_D} dx \frac{x^4 e^x}{(e^x-1)^2} + 3R\left(\frac{\theta_E}{T}\right)^2 \frac{e^{\frac{\theta_E}{T}}}{\left(e^{\frac{\theta_E}{T}}-1\right)^2}$$

Where R is the gas constant, $\theta_D$ is the Debye temperature, $x_D = \left(\frac{\theta_D}{T}\right)$ and $\theta_E$ is the Einstein Temperature

For fitting both Debye terms and Einstein terms were taken. For instance, Debye Einstein fit of $In_2Te_5$ shown inside the inset fig 3(b). For $In_2(Te_{1-x}Se_x)_5$ x = 0, 0.05, 0.10 and $Fe_{0.05}In_{1.95}(Te_{0.90}Se_{0.10})_5$ samples corresponding values of $\theta_D$s are 231, 268, 215 and 234 K respectively and $\theta_E$s are 57, 61, 58 and 52 K respectively [18, 19]. The calculated sound velocities are obtained by the relationship of $\theta_D = v_s\left(\frac{\hbar}{k_B}\right)(6\pi^2 n)^{\frac{1}{3}}$ [18] where n is number density. Using the phonon heat capacities, sound velocities and lattice thermal conductivities, an estimation of phonon mean free paths was done which is shown in fig 3(b). Increase the Se concentration up to x = 0.10, in $In_2(Te_{0.90}Se_{0.10})_5$ increases the $l_{ph}$ up to 1.73 µm and for $Fe_{0.05}In_{1.95}(Te_{0.90}Se_{0.10})_5$ shows minimum $l_{ph}$ of about 0.79 µm about 195 K. It reveals that through the Fe substitution, an enhanced phonon scattering process along with a suppressed phonon mean free path (which also depend on the large atomic mass difference between Fe and In) results. The inset in 3(b) shows the measured temperature dependent heat capacity. Observed lower heat capacity for Fe doped sample as compare parent compound and for $In_2(Te_{0.90}Se_{0.10})_5$ $x = 0.10$ shows higher heat capacity as compare to other samples.

Figure 4(a) shows the power factors for $In_2(Te_{1-x}Se_x)_5$ $x = 0, 0.05, 0.10$ and $Fe_{0.05}In_{1.95}(Te_{0.90}Se_{0.10})_5$. It is observed that increasing with Se doping, the power factor is enhanced. The highest value of power factor is 72 µW/mK² at 347 K in $Fe_{0.05}In_{1.95}(Te_{0.90}Se_{0.10})_5$ sample. An enhancement in power factor is mainly due to the reduced electrical resistivity.

Figure 4(b) shows the temperature dependence of $zT$ for $In_2(Te_{1-x}Se_x)_5$ $x = 0, 0.05, 0.10$ and $Fe_{0.05}In_{1.95}(Te_{0.90}Se_{0.10})_5$. $zT$ increases with temperature, with maximum $zT$ value $\approx 3.7 \times 10^{-3}$ and $\approx 7.6 \times 10^{-3}$ around 260 K for $In_2(Te_{0.95}Se_{0.05})_5$ $x = 0.05, 0.10$ samples which is 9.25 and 18.93 times larger than the parent compound ($ZT \approx 4.0 \times 10^{-4}$) respectively. This mainly due to improved electrical conductivity and enhanced Seebeck. While for $Fe_{0.05}In_{1.95}(Te_{0.90}Se_{0.10})_5$ sample $zT$ achieved maximum value $\approx 0.076$ which is about 310 times higher than the $In_2Te_5$ ($zT \approx 2.45 \times 10^{-4}$) around 347 K. That much enhanced $zT$ is due mainly to reduced thermal conductivity and reduced electrical resistivity. The value of $zT$ could be further increased by optimizing Fe concentration [15].

4. Conclusions

In summary, we have successfully synthesized and characterized p - type thermoelectric properties of $In_2(Te_{1-x}Se_x)_5$ where $x = 0, 0.05, 0.10$ and $Fe_{0.05}In_{1.95}(Te_{0.90}Se_{0.10})_5$ samples. At 10% Se substitution, concentration reduced resistivity up to 0.0065 Ωm and increased S of 535μV/K at about 219 K. For $Fe_{0.05}In_{1.95}(Te_{0.90}Se_{0.10})_5$ lattice thermal conductivity and resistivity is decreased to 0.31 W/K-m and 0.00039 Ωm at 330 K respectively. Even though the value of S deteriorates there is an overall improvement in *zT*. The major reduction in the lattice thermal conductivity is due to the large atomic mass difference between dopant Fe and host In [15]. The obtained maximum *zT* value is 0.076 at 347 K for $Fe_{0.05}In_{1.95}(Te_{0.90}Se_{0.10})_5$ sample.

**Acknowledgments**

Authors would like to acknowledge the Indian Department of Science and Technology for partial support through the project IR/S2/PU-10/2006. ADT would like to acknowledge partial support from the Center for Energy and Environment, IIT Patna.

**References**

[1] J. R. Sootsman, D. Y. Chung, and M. G. Kanatzidis, Angew. Chem. Int. Ed. **48**, 8616 (2009).

[2] B. Yang, H. Ahuja, and T. N. Tran, HVAC & R RESEARCH **14**, 635 (2008).

[3] G. S. Nolas, J. Sharp, and H. J. Goldsmid, Thermoelectrics: Basic Principles and New Materials Developments (Springer, 2001).

[4] C. Wood, Rep. Prog. Phys. **51**, 459 (1988).

[5] T. M. Tritt, M. Kanatzidis, G. Mahan & H. B. Lyon, Mater. Res. Soc. Symp. Proc. **478**, (1997).

[6] T. Hendricks, and W. T. Choate, Engineering Scoping Study of Thermoelectric Generator Systems For Industrial Waste Heat Recovery, (US Department of Energy: 2006).

[7] J. Yang, Potential applications of thermoelectric waste heat recovery in the auto-motive industry, Proceedings of 24th International Conference on Thermoelectrics, IEEE Catalog No. 0-7803-9552-2/05, P. 155, (2005).

[8] X. Shi, J. Y. Cho, J. R. Salvador, J. Yang, H. Wang, Appl. Phys. Lett. **96,** 162108 (2010).

[9] G. H. Zhu, Y. C. Lan, H. Wang, G. Joshi, Q. Hao, G. Chen, and Z. F. Ren, Phys. Rev. B. **83**, 115201 (2011).

[10] J. S. Rhyee, K. H. Lee, S. M. Lee, E. Cho, S. Kim, E. Lee, Y. S. Kwon, J. H. Shim, and G. Kotliar, *Nature* **459**, 965 (2009).

[11] J. S. Rhyee, and J. H. Kim, Materials **8**, 1283 (2015).

[12] L. D. Zhao, S. H. Lo, Y. Zhang, H. Sun, G. Tan, C. Uher, C. Wolverton, V. P. Dravid & G. Kanatzidis, Nature **508**, 303 (2014).

[13] M. M. Nassary, M. Dongal, M. K. Gerges & M. A. Sebage, Phys. Stat. Sol.(a) **199**, 464 (2003).

[14] A. V. Sanchela, A. D. Thakur & C. V. Tomy AIP Conf. Proc. **1591**, 1392 (2014)

[15] H. Mun, K. H. Lee, S. J. Kim, J. Y. Kim, J. H. Lee, J. H. Lim, H. J. Park, J. W. Roh, and S. W. Kim, Materials **8**, 959 (2015)

[16] X. Liang, ACS Appl. Mater. Interfaces **7**, 7927 (2015).

[17] T. M. Tritt, Thermal Conductivity, Theory, Properties and Applications (Kluwer Academic / Plenum Publishers 2004).

[18] C. Kittel, Introduction to Solid State Physics (7$^{th}$ edition, John Wiley, 1996).

[19] Y. Du, Z. Cheng, S. Dou, X. Wang, and H. Zhao, Appl. Phys. Lett. **97**, 122502 (2010).

**Figure Captions**

**Figure 1** (Color online) **(a)** X-ray diffraction pattern for powdered single crystalline sample of $In_2Te_5$ with Rietveld refinement was performed using the monoclinic structure; space group C12/c1, (PDF # 00-031-0602) and R factors $R_p$ = 6.9 %, $R_{WP}$ = 9.2 %. **Figure 1(b), (c), (d)** and **(e)** shows XRD taken of single crystals with proper alignment and indicates crystals grown along the c axis.

**Figure 2** (Color online) **(a)** Resistivity $\rho$ and **(b)** Seebeck coefficient $S$ as a function of temperature for $In_2(Te_{1-x}Se_x)_5$ x = 0, 0.05, 0.10 and $Fe_{0.05}In_{1.95}(Te_{0.90}Se_{0.10})_5$.

**Figure 3** (Color online) **(a)** Lattice thermal conductivity $\kappa_{lat}$ (main panel), total thermal conductivity $\kappa_{tot}$ (left inset) and electronic thermal conductivity $\kappa_e$ (right inset), **(b)** Phonon mean free path $l_e$ and heat capacity $C_p$ (inset) as a function of temperature for $In_2(Te_{1-x}Se_x)_5$ x = 0, 0.05, 0.10 and $Fe_{0.05}In_{1.95}(Te_{0.90}Se_{0.10})_5$. For instance Debye Einstein fitting shows for $In_2Te_5$ inside the inset fig 3(b).

**Figure 4** (Color online) **(a)** Power Factor *PF* and **(b)** Figure of merit *zT* as a function of temperature for $In_2(Te_{1-x}Se_x)_5$ x = 0, 0.05, 0.10 and $Fe_{0.05}In_{1.95}(Te_{0.90}Se_{0.10})_5$.

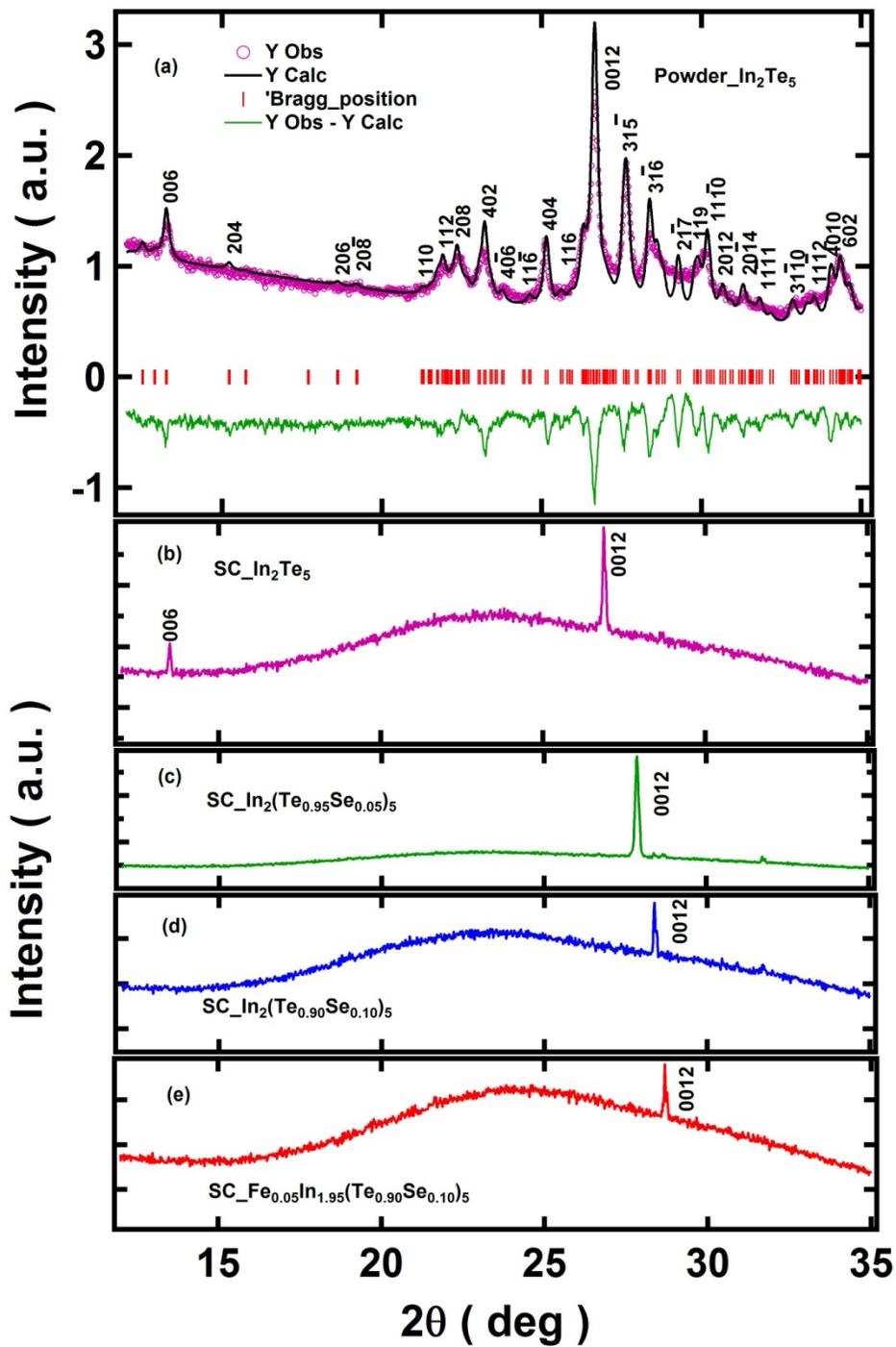

**FIGURE 1 (a), (b), (c), (d) and (e)**

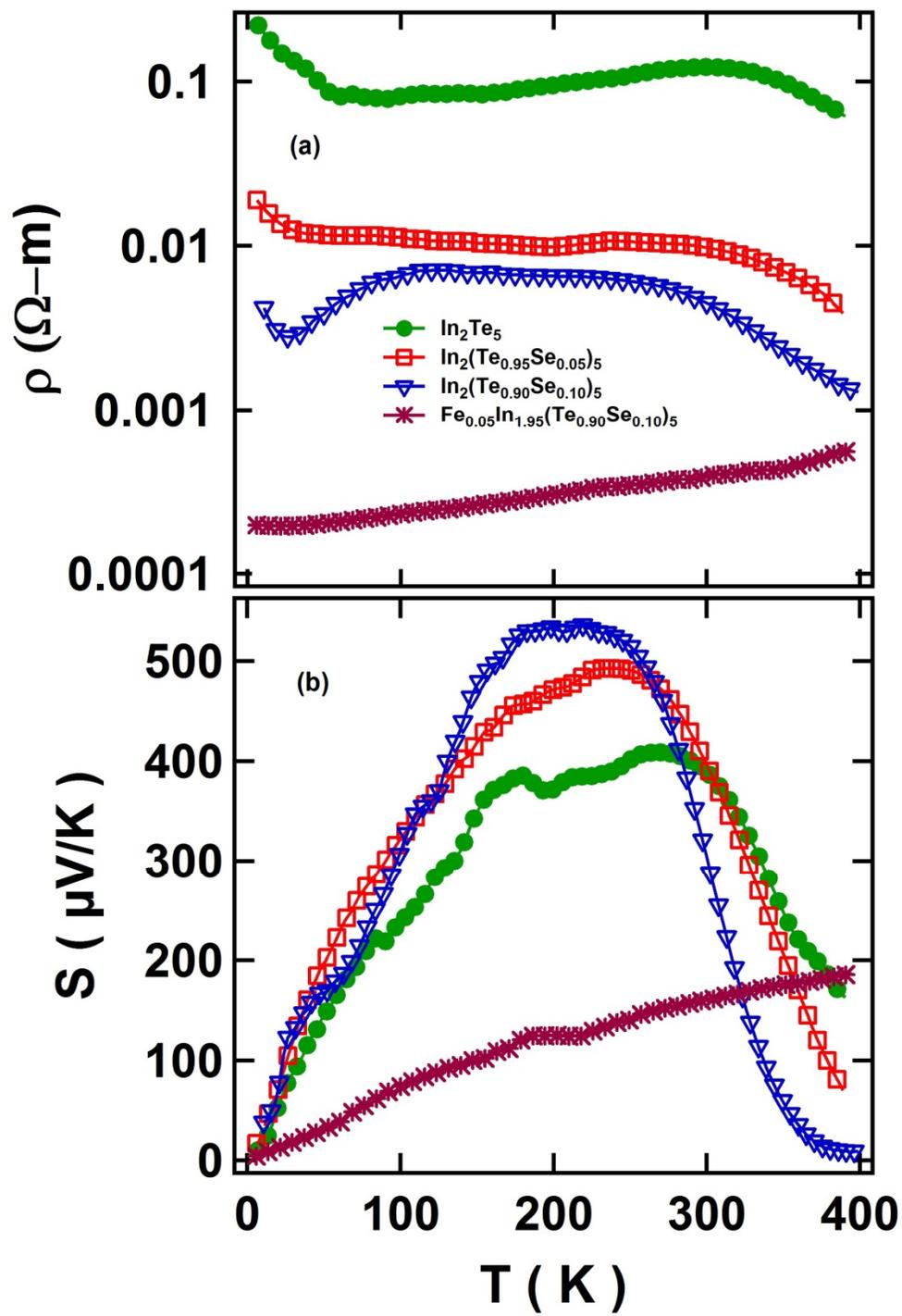

**FIGURE 2** (a), (b)

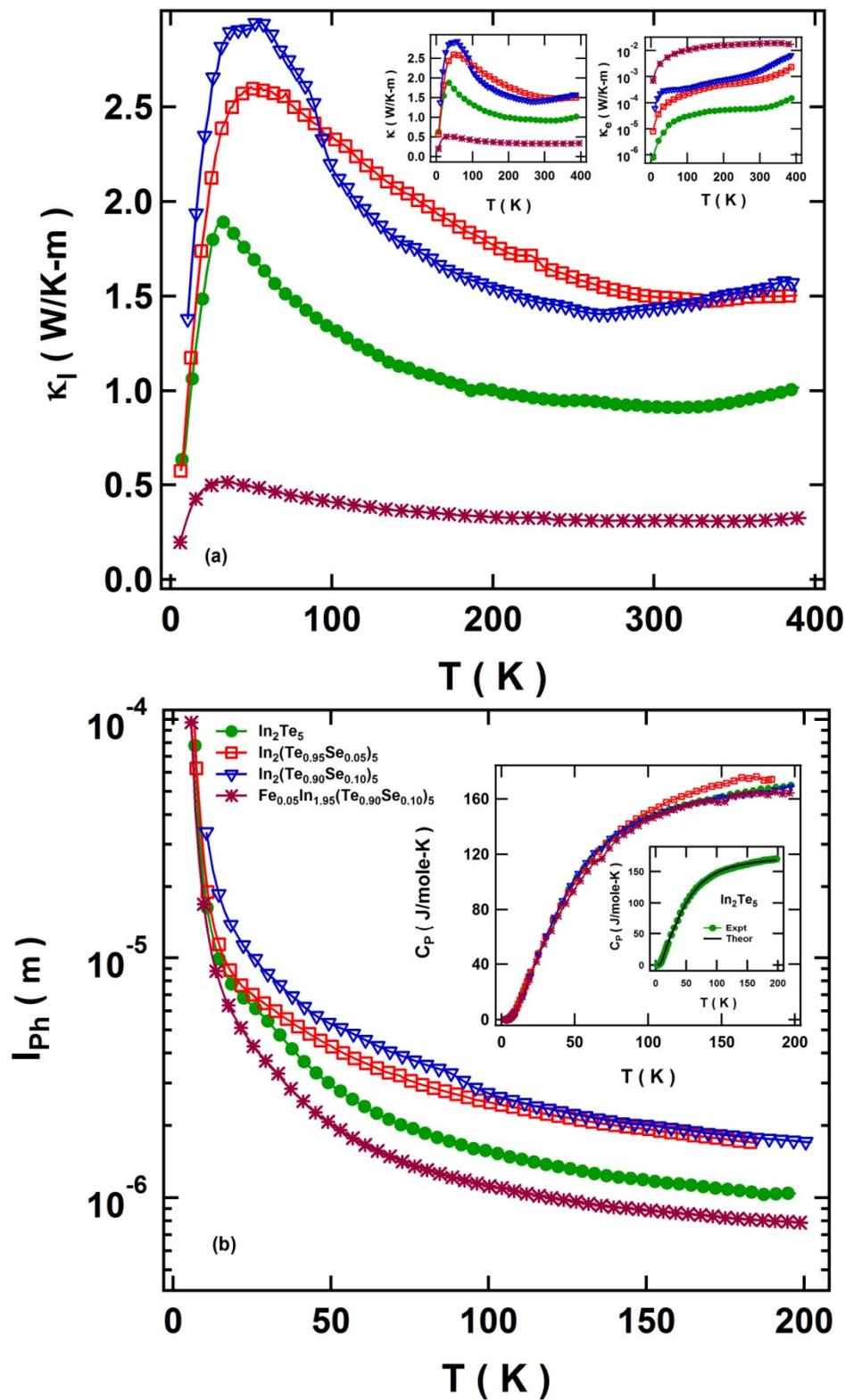

**FIGURE 3 (a), (b)**

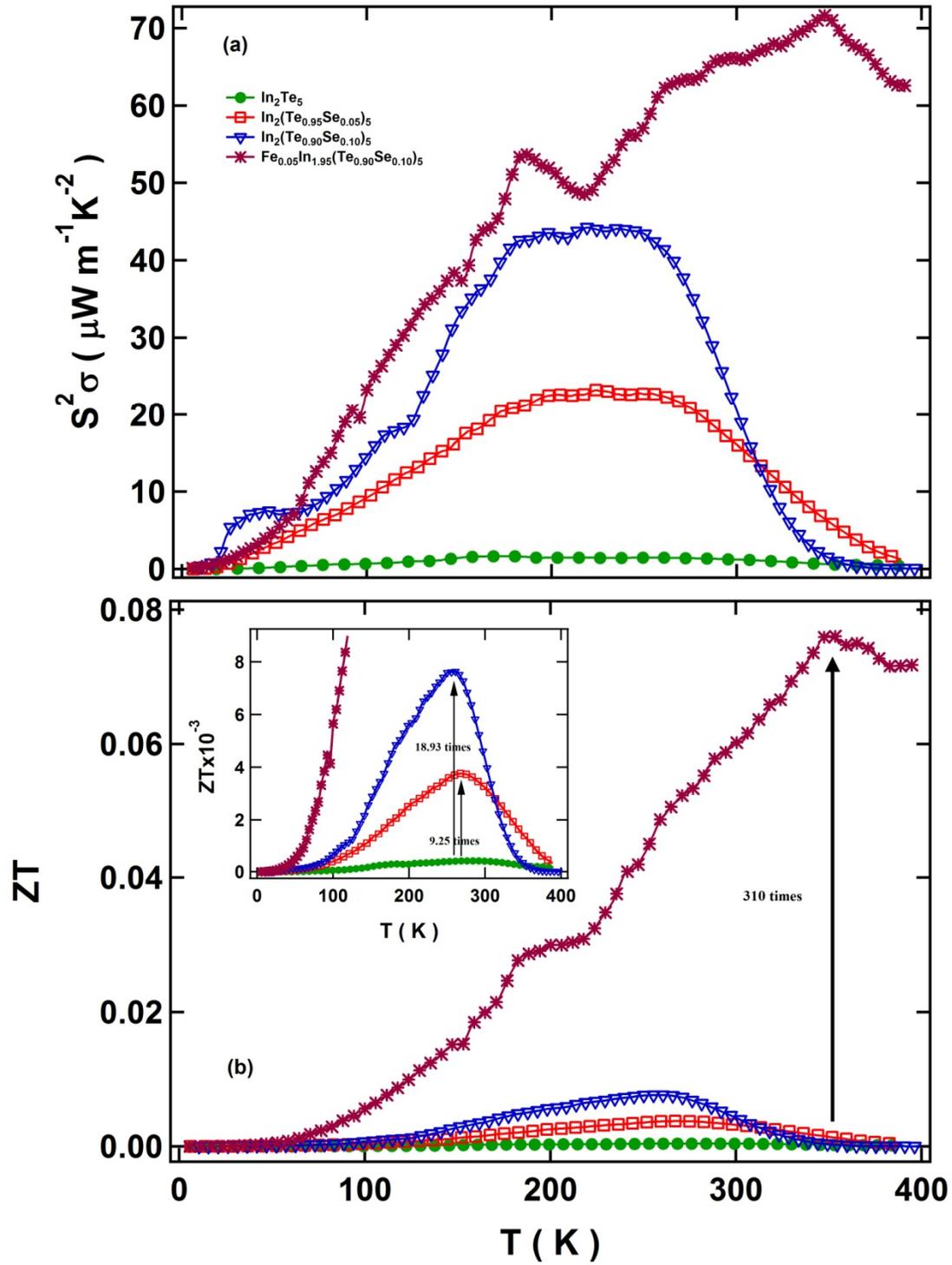

**FIGURE 4 (a), (b)**